\documentclass[final,5p,times,twocolumn]{elsarticle}
\usepackage{graphicx}
\usepackage{amssymb}
\usepackage{subeqn}

\begin{document}
\title{Constraining Decaying Dark Matter}
\author{Ran Huo}
\address{Enrico Fermi Institute, University of Chicago, Chicago, Il 60637, United States}

\begin{abstract}
We revisited the decaying dark matter (DDM) model, in which one collisionless particle decays early into two collisionless particles, that are potentially dark matter particles today. The effect of DDM will be manifested in the cosmic microwave background (CMB) and structure formation. With a systematic modification of CMB calculation tool \texttt{camb}, we can numerically calculated this effect, and compare it to observations. Further Markov Chain Monte Carlo \texttt{cosmomc} runnings update the constraints in that model: the free streaming length $\lambda_{\mbox{{\scriptsize FS}}}\lesssim0.5$Mpc for nonrelativistic decay, and $(\frac{M_{\mbox{{\tiny DDM}}}}{\mbox{{\scriptsize keV}}}Y)^2\frac{T_d}{\mbox{{\scriptsize yr}}}\lesssim5\times10^{-5}$ for relativistic decay.
\end{abstract}
\maketitle

\section{Introduction}

Decaying dark matter (DDM) in which the decay process happens at an early stage of the universe is natural for many models, and it is sometimes introduced as a way to adjust the DM relic density, because after decay the DM relic density will naively be lowered by
\begin{equation}\label{Omega}
\Omega_{d}=\frac{\sum m_{\mbox{{\tiny product}}}}{m_{\mbox{{\tiny original}}}}\Omega_{\mbox{{\scriptsize DDM}}}.
\end{equation}
One example is in \cite{Asaka:2000ew}, in which gravitino, overproduced by reheating after inflation, will decay into the true lightest supersymmetric particles (LSP) axino as well as an axion.

However, pure gravitational constraints for that decay process are less understood and sometimes even simply ignored. Actually the model in \cite{Asaka:2000ew} with their parameters should be ruled out \cite{Huo:2010ph}. Here we will present a model independent computation, in which the effect is calculated from the first principle, and can be compared directly with cosmological observation. Our model have both the parent particle and the daughter particles interacting very weakly, so the effect can only be manifested in gravitational effect, such as the cosmic microwave background (CMB) and the large scale structure formation. Our approach is based on a systematic modification of the CMB codes.

\section{Principle}

\texttt{cmbfast}\cite{Seljak:1996is} and \texttt{camb}\cite{Lewis:1999bs} are CMB calculation tools which are based on the photon line of sight integration technique, instead of solving the Boltzmann equation explicitly. They work in the \emph{synchronous gauge}, the metric perturbation of which is gauged completely into the spatial $3\times3$ part
\begin{equation}
g_{\mu\nu}=a^2\left(\begin{array}{cc}
-1 & \\
& \delta_{ij}+h_{ij}
\end{array}\right),
\end{equation}
and the graviton $h_{ij}$ can be decomposed into Fourier modes $h$ and $\eta$
\begin{equation}
h_{ij}(\vec{x})=\int\frac{d^3k}{(2\pi)^{\frac{3}{2}}}e^{i\vec{k}\cdot\vec{x}}\bigg(\hat{k}_i\hat{k}_jh(\vec{k})+(\hat{k}_i\hat{k}_j-\frac{1}{3}\delta_{ij})6\eta(\vec{k})\bigg).
\end{equation}

Then linearized Einstein equation gives the equations of motion of $h$ and $\eta$ components \cite{Ma:1995ey}
\begin{subeqnarray}\label{handeta}
k^2\eta-\frac{1}{2}\frac{\dot{a}}{a}\dot{h}&=&4\pi Ga^2\delta T^0_0,\\
k^2\dot{\eta}&=&4\pi Ga^2(\bar{\rho}+\bar{p})\theta,\\
\ddot{h}+2\frac{\dot{a}}{a}\dot{h}-2k^2\eta&=&-8\pi Ga^2\delta T^i_i,\\
\ddot{h}+6\ddot{\eta}+2\frac{\dot{a}}{a}(\dot{h}+6\dot{\eta})-2k^2\eta&=&-24\pi Ga^2(\bar{\rho}+\bar{p})\sigma.
\end{subeqnarray}
Here overdot $\dot{\enspace}$ means derivative to conformal time $\tau$. $\theta$ is the peculiar velocity which is defined by $(\bar{\rho}+\bar{p})\theta\equiv ik^i\delta T^0_i$, and $\sigma$ is the shear which is defined by $(\bar{\rho}+\bar{p})\sigma\equiv -(\hat{k}_i\hat{k}_j-\frac{1}{3}\delta_{ij})\Sigma^i_j=-(\hat{k}_i\hat{k}_j-\frac{1}{3}\delta_{ij})(T^i_j-\frac{1}{3}\delta^i_jT^k_k)$.

With the $h$ and $\eta$ metric perturbation, the Boltzmann equation in terms of the fractional perturbation $\Psi$ is \cite{Ma:1995ey}
\begin{eqnarray}\label{Boltzmann}
\frac{\partial\Psi}{\partial\tau}+i\frac{qk}{\epsilon}(\hat{k}\cdot\hat{n})\Psi+\frac{\partial\ln f(q)}{\partial\ln q}\bigg(\dot{\eta}-\frac{\dot{h}+6\dot{\eta}}{2}(\hat{k}\cdot\hat{n})^2\bigg)\nonumber\\
=\frac{1}{f(q)}\bigg(\frac{\partial(f+\delta
f)}{\partial\tau}\bigg)_{C},
\end{eqnarray}
where $q=ap$ is the comoving momentum and is \emph{conserved} in expansion if the particle is collisionless, $\epsilon=\sqrt{q^2+a^2m^2}$ is the comoving energy, and $\hat{n}$ is the direction of the macroscopic flow of the fluid. $f(q)$ is the unperturbed partition function, and the real partition function can be defined with fractional perturbation $f(q,\tau,x^i,n_i)=f(q)\Big(1+\Psi(q,\tau,x^i,n_i)\Big)$. Usually $f(q)$ is thermal distribution such as Fermi-Dirac distribution or Bose-Einstein distribution, but in our DDM model for daughter particles it is determined by the decay process.

After we plug into the collision term, we will find the formal photon line-of-sight integration solution to the Boltzmann equation. The anisotropy $\Delta_T\equiv\frac{\delta T}{\bar{T}}\sim\frac{1}{4}\Psi_\gamma$ \emph{today} is given by \cite{Seljak:1996is}
\begin{eqnarray}\label{lineofsight}
\Delta_T(\vec{k},\hat{n})&=&\int_0^{\tau_0}d\tau
e^{ik\mu(\tau-\tau_0)}e^\kappa\Big[(\dot{\eta}-\alpha\mu^2k^2)\nonumber\\
&\hspace{-6em}+&\hspace{-3.2em}\dot{\kappa}\big[\Delta_{T0}+\mu
v_e-\frac{1}{2}P_2(\mu)(\Delta_{T2}+\Delta_{P0}+\Delta_{P2})\big]\Big].
\end{eqnarray}
where the optical depth is $\kappa\equiv-\int^{\tau_0}_\tau d\tau'\dot{\kappa}(\tau')<0$ and the differential optical depth is $\dot{\kappa}=an_e\sigma_T$, which is the common factor for all collision terms. Here $n_e$ is the number density of \emph{free} electrons in coordinate space and $\sigma_T=6.65\times10^{-25}$ cm$^2$ is the Thomson cross section. $\mu\equiv(\hat{k}\cdot\hat{n})$, $\alpha\equiv\frac{\dot{h}+6\dot{\eta}}{2k^2}$ and $v_e$ is the electron velocity. Suffix P means polarization mode and integer suffix labels multipoles of spherical harmonics.

In DDM model, metric perturbations $h$ and $\eta$ get significant contribution from the daughter particles. The perturbation evolution of the daughter particle has the same series with the massive neutrino
\begin{subeqnarray}\label{massive neutrino}
\dot{\Psi}_0&=&-\frac{qk}{\epsilon}\Psi_1+\frac{1}{6}\dot{h}\frac{d\ln f}{d\ln q},\\
\dot{\Psi}_1&=&\frac{qk}{\epsilon}\left(\Psi_0-\frac{2}{3}\Psi_2\right),\\
\dot{\Psi}_2&=&\frac{qk}{\epsilon}\left(\frac{2}{5}\Psi_1-\frac{3}{5}\Psi_3\right)-\left(\frac{1}{15}\dot{h}+\frac{2}{5}\dot{\eta}\right)\frac{d\ln f}{d\ln q},\\
\dot{\Psi}_\ell&=&\frac{qk}{\epsilon}\frac{1}{2\ell+1}\bigg(\ell\Psi_{\ell-1}-(\ell+1)\Psi_{\ell+1}\bigg)\quad\ell\geq3,\\
\dot{\Psi}_\ell&=&\frac{qk}{\epsilon}\Psi_{\ell-1}+\frac{\ell+1}{\tau}\Psi_\ell\qquad\mbox{As Truncation}.
\end{subeqnarray}
The right hand side source term of Eq.\ (\ref{handeta}) have contributions only from the first three perturbation modes ($\delta\rho,\thinspace\delta p\propto\Psi_0$, $\theta\propto\Psi_1$ and $\sigma\propto\Psi_2$). Because all particle species such as baryon, cold dark matter (CDM), photon, massless as well as massive neutrino talk to gravity, all their evolution will be modified.

The codes also calculate the transfer functions (TF) as an intermediate step in the CMB calculations, which is an indication of the large scale structure. The TF is by definition the normalized ratio of perturbation growth factor, from an very early stage to certain late stage, the normalization is taken with a very large scale which is out of horizon in the whole evolution
\begin{equation}
T(k)\equiv\frac{{\displaystyle \delta(k,t_f)/\delta(k,t_i)}}{{\displaystyle \delta(k\rightarrow0,t_f)/\delta(k\rightarrow0,t_i)}}.
\end{equation}

\section{Modification}

We introduce free parameter $\Omega_d$ which corresponds to the daughter particles' energy density \emph{today}, while still keeping the nondecay CDM part $\Omega_c$. In this way we can treat any combination of decaying and nondecay dark matter. For convenience to use Eq.\ (\ref{Omega}), $\Omega_d$ is not the whole energy density but only the mass contribution to energy density, namely the kinetic energy of the daughter particle is not included. Therefore we should have at least one massive daughter particle for this parameterization. Except for an extreme relativistic decay, the kinetic energy contribution is small and $\Omega_d$ represents the energy density very well. We only consider the one-to-two decay process and introduce two mass ratios $\frac{m_{p1}}{m_o}$ and $\frac{m_{p2}}{m_o}$, where $m_{p1}$ and $m_{p2}$ are separately the masses of two product particles and $m_{o}$ is the mass of original particle. The last free parameter is the decay lifetime $T_d$. So the complete set of new parameters includes $\Omega_d$, $T_d$, $\frac{m_{p1}}{m_o}$ and $\frac{m_{p2}}{m_o}$.

Let us go through what will be modified in our DDM model, compared with the standard $\Lambda$CDM universe. First, the decay process will affect the expansion of the universe, through changing the equation of state. Before decay the DDM behaves as the CDM with a constant equation of state $\omega=0$, while after decay it does not hold and the momentum and energy of daughter particle is subject to redshift, as what happens to massive neutrino. Since before decay the DDM particle can be approximated as being at rest, given the masses of the parent and daughter particles the initial transverse momentum $p_T$ is fixed for a two body decay. The comoving momentum for each individual daughter particle is determined only by the scale factor $a^\ast$ at which the decay happens (in this paper we will always use a $^\ast$ to denote the quantity right at decay). As the scale factor $a(\tau)$ grows the physical momentum $p(\tau)$ decreases, while preserving
\begin{equation}
q=a(\tau)p(\tau)=a^\ast p_T.
\end{equation}
With this relation the energy density and pressure can be evaluated numerically for product particles in the modification, so is the expansion process.

The decay is a continuous process for the set of DDM particles, the way for our numerical study is to discretize it into $30\sim60$ channels, by which we can achieve 0.5\% precision for CMB peak height. Each channel corresponds to daughter particles produced in a small scale factor region $a^\ast\sim a^\ast+da^\ast$ and has its own $\Psi_\ell$ series. As the universe expands the channels are gradually filled channel by channel in order. The perturbation evolution is described by Eq.\ (\ref{massive neutrino}), the only subtlety comes through the factor $\frac{d\ln f}{d\ln q}$: the unperturbed distribution $f(q)$ is no longer the Fermi-Dirac distribution of neutrino, but determined by the decay process. A number of product particles proportional to $d\Omega$ will be redistributed into $q$ space $d^3q=4\pi q^2dq=4\pi p_T^3a^{\ast2}da^\ast$, so up to some factor the unperturbed partition function from decay is
\begin{equation}
f\equiv\frac{dn}{d^3p}\propto\frac{d\Omega}{d^3q}=\frac{e^{-\frac{t}{T_d}}\frac{dt}{T_d}}{4\pi p_T^3a^{\ast2}da^\ast}=\frac{e^{-\frac{t}{T_d}}}{4\pi T_dp_T^3a^\ast\dot{a}^\ast},
\end{equation}
and $\frac{d\ln f}{d\ln q}$ can be calculated
\begin{equation}\label{dlnfdlnq}
\frac{d\ln f}{d\ln q}=-\frac{a^{\ast2}}{T_d\thinspace\dot{a}^\ast}-\frac{3}{2}+\frac{3\bar{p}^\ast}{2\bar{\rho}^\ast}.
\end{equation}

The last issue for the perturbation evolution differential equation set is the initial condition. The initial values of all perturbation modes should naturally inherit the values before decay, which for DDM they are the same as CDM and all higher multipoles vanish. So we have
\begin{eqnarray}
&&\Psi_0^\ast=\delta_{\mbox{{\scriptsize CDM}}},\\
&&\Psi_\ell^\ast=0 \qquad\ell\geq1.
\end{eqnarray}

Including the daughter particles' contribution to metric perturbation finishes our modification.

\section{Sample Calculation}

Generally speaking, because DDM induces no modification of the photon-electron-baryon plasma, the CMB anisotropy spectrum is only directly affected by the Sachs-Wolfe (SW) effect through metric perturbation. Relativistic particles contribute more than massive particles to the peculiar velocity and the shear, because they are not suppressed by $\frac{v}{c}$ or $\frac{v^2}{c^2}$ factors. So the SW effect will be enhanced dominantly by the relativistic decay products, and the acoustic peaks will raise and move to higher multipoles $\ell$, as what happens to a common SW effect.

As for the matter power spectrum, structure formation requires DM particles to condensate into clumps, in order to amplify the density perturbation. Therefore the DM particles must be moving slowly to be gravitationally captured, just like a slow incident comet will be captured by the sun to form an elliptic orbit but not the fast moving one with a hyperbolic orbit. If DDM is dominant and nondecay CDM doesn't exist or is negligible, after decay on small scale the free streaming effect will prevent structures from growing, and the TF will be much lower than the nondecay case. If there is still sizable CDM, the small scale power is not completely erased, but only get smaller.

For certain decay lifetime between nucleosynthesis and recombination, there are two different approaches which may make the DDM model work. One is that the decay may happens to the dominant part of the dark matter, but the daughter particle has mass close to the parent particle and is very nonrelativistic, at least after redshift at the late recombination and structure formation epoch, so that the decay effect is minimized. The other is that the daughter particle is light and can be all the way relativistic, but only a tiny part of the dark matter today comes from it, so that its effect is constrained. We will separately call them nonrelativistic scenario and relativistic scenario. A combination of the two works for sure, but their individual effects are primarily interesting.

\begin{figure}
\includegraphics[height=2.8in]{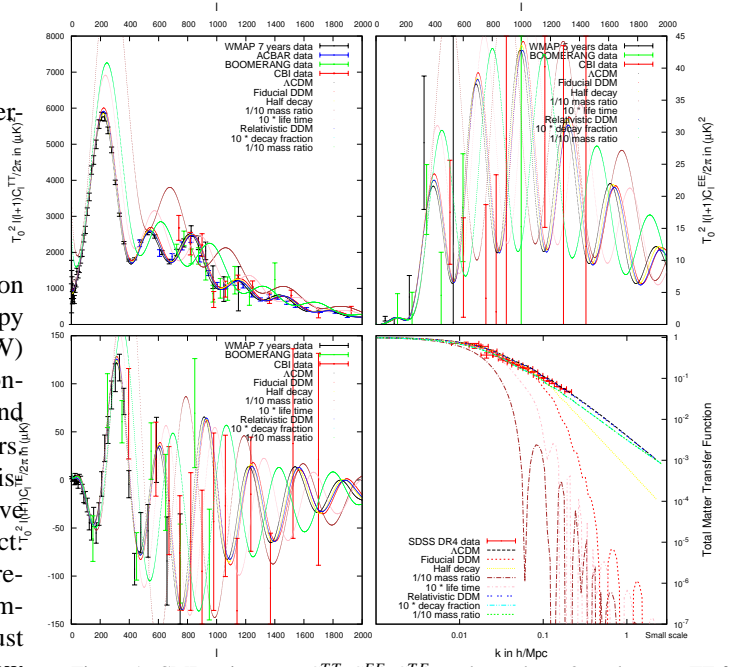}
\caption{\label{sample}CMB anisotropy $C_\ell^{TT}$ $C_\ell^{EE}$ $C_\ell^{TE}$ modes and $z=0$ total matter TF for various DDM model. The ``$\Lambda$CDM'' model (black) has the best fit cosmological parameters which apply implicitly to the following models. The ``Fiducial DDM'' model (red) has $\Omega_d=0.226$ completely replacing CDM, two mass ratios $\frac{m_{p1}}{m_o}=0.1$ and $\frac{m_{p2}}{m_o}=0$, and decay lifetime $T_d=10^9$s. Then we separately vary $\Omega_d$ to be half of the fiducial value while the other half is still the CDM (yellow), mass ratio $\frac{m_{p1}}{m_o}=0.01$ (brown) and decay lifetime $T_d=10^{10}$s (pink). We also show an example of very ``Relativistic DDM'' model (blue), which has $\Omega_d=10^{-5}$ and CDM component the same as the $\Lambda$CDM model, two mass ratios $\frac{m_{p1}}{m_o}=10^{-5}$ (so that $\Omega_{\mbox{{\scriptsize DDM}}}=1$) and $\frac{m_{p2}}{m_o}=0$, and decay lifetime $T_d=10^9$s. We then vary $\Omega_d$ to be $10^{-4}$ (cyan) and $\frac{m_{p1}}{m_o}$ to be $10^{-6}$ (green).}
\end{figure}
We present our sample calculations in Fig.\ \ref{sample}. As expected, as we go to more DDM compoenent, larger mass hierarchy and longer decay lifetime, we see greater SW effect and greater free streaming effect. The TF is also calculated in \cite{Kaplinghat:2005sy}. In models with no CDM we see very suppressed TF on small scales, and the oscillation structure of which is given by baryon acoustic oscillation. The two variants of ``Relativistic DDM'' model coincide nearly exactly, which implies in that limit only $\Omega_{\mbox{{\scriptsize DDM}}}$ matters.

\section{Parameter Constraint}

We have further run our modified \texttt{camb} as calculation tool for the Markov Chain Monte Carlo (MCMC) program \texttt{cosmomc} \cite{Lewis:2002ah} for cosmological parameter evaluation. We are equipped with all current public data except for the SDSS DR8: 4 CMB data sets which are WMAP 7 years, ACBAR, CBI and BOOMERANG \cite{Komatsu:2010fb}; 3 matter power spectrum data sets which are SDSS DR7 LRG, SDSS DR4 and 2dFGRS \cite{Reid:2009xm}, as well as supernova data. We are constrained to minimal $\Lambda$CDM plus DDM model, which is a flat universe of $\Omega_k=0$, no hot dark matter $\Omega_\nu=0$, standard cosmological constant $w=-1$, no tensor mode $r=0$ and no spectral index running $n_{\mbox{{\scriptsize run}}}=0$. The channels we used are CMB, HST, mpk, BBN, Age Tophat Prior and SN, which are consistent with that model and independent from each other. In addition to the 4 decay parameters there are 7 other parameters subject to Monte Carlo: $\Omega_bh^2$, $\Omega_ch^2$, $\theta$, $z_{\mbox{{\scriptsize rei}}}$, $n_s$, $\log A$, $A_{SZ}$.

\begin{figure}
\includegraphics[height=1.8in]{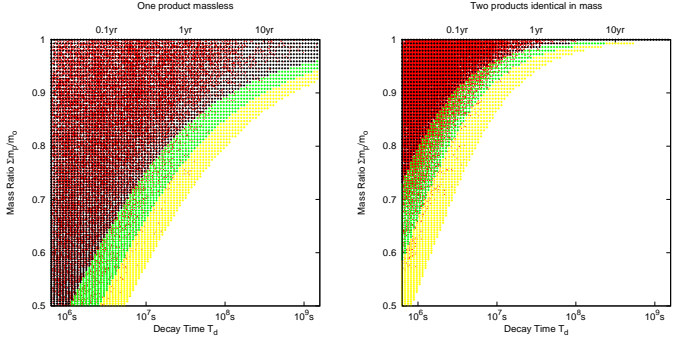}
\caption{\label{nonrelativisticplot}Decay lifetime and mass ratio regions for nonrelativistic decay scenario, where the DM is pure DDM and one product is massless (left) or two products have identical masses (right). The black gridded regions which contain 95\% MCMC points (red dots) correspond to $\lambda_{\mbox{{\scriptsize FS}}}\leq0.49$Mpc (left) or $\lambda_{\mbox{{\scriptsize FS}}}\leq0.50$Mpc (right), and the green gridded regions which contain 99.7\% points correspond to $\lambda_{\mbox{{\scriptsize FS}}}\leq0.72$Mpc (left) or $\lambda_{\mbox{{\scriptsize FS}}}\leq0.73$Mpc (right), comparing to the yellow gridded regions of previous constraint $\lambda_{\mbox{{\scriptsize FS}}}\leq1$Mpc \cite{Cembranos:2005us}.}
\end{figure}
Although a running for more general parameter space is possible, limited by our computing facility we are still working in the nonrelativistic scenario and the relativistic scenario, without exploring the intermediate region. First we consider the nonrelativistic decay scenario. We further constrain that all dark matter undergoes decay, which trade $\Omega_d$ into $\Omega_c$. Then we focus on some certain relation between two daughter particles' mass for further simplification. Corresponding to \cite{Asaka:2000ew} we do the case that one product is massless, the other one which we focused on is that the two products have the same mass. The results are shown in Fig.\ \ref{nonrelativisticplot}.

We find the allowed region can be described concisely by the free streaming length constraint $\lambda_{\mbox{{\scriptsize FS}}}\lesssim0.5$Mpc at 95\% confidence level. Here the free streaming length is defined as the length measured today, where a DDM particle decays exactly at its expectation lifetime $T_d$, and the daughter particle travels before matter-radiation equality. Normalized to $a_0=1$, we have
\begin{equation}
\lambda_{\mbox{{\scriptsize FS}}}\equiv\int_{\tau_d}^{\tau_{\mbox{\tiny eq}}}\hspace{-0.3em}v(\tau)d\tau=\int_{a_d}^{a_{\mbox{\tiny eq}}}\hspace{-0.3em}\frac{q}{\sqrt{m^2a^2\hspace{-0.2em}+q^2}}\frac{da}{\sqrt{\frac{8\pi G}{3}(\rho_{m0}a+\rho_{r0})}}.
\end{equation}
In the identical products' masses case the free streaming length is the same for the two particles; and in the one massless product particle case it is defined for the massive particle, since it contributes to the structure formation. In fact we also checked the intermediate case such as one daughter particle is three times heavier than the other, in which the free streaming length cannot be defined without ambiguity, and the numerical allowed region is a reasonable average of the two particles' regions corresponding to the same free streaming length at the same confidence level.

\begin{figure}
\includegraphics[height=1.8in]{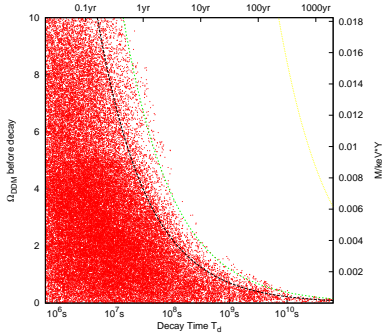}
\caption{\label{relativisticplot}Decay lifetime and $\Omega_{\mbox{{\scriptsize DDM}}}$ region for relativistic decay scenario. Here we take $h^2=0.5$. The black line which contains 95\% MCMC points (red dots) is the constraint $(\frac{M_{\mbox{{\tiny DDM}}}}{\mbox{{\scriptsize keV}}}Y)^2\frac{T_d}{\mbox{{\scriptsize yr}}}\leq5.3\times10^{-5}$, and the green line which contains 99.7\% points is $(\frac{M_{\mbox{{\tiny DDM}}}}{\mbox{{\scriptsize keV}}}Y)^2\frac{T_d}{\mbox{{\scriptsize yr}}}\leq1.5\times10^{-4}$, comparing to the yellow line of previous constraint $(\frac{M_{\mbox{{\tiny DDM}}}}{\mbox{{\scriptsize keV}}}Y)^2\frac{T_d}{\mbox{{\scriptsize yr}}}\leq7.5\times10^{-2}$ \cite{Zentner:2001zr}.}
\end{figure}

Then we consider the relativistic decay scenario. Here we use Eq.\ (\ref{Omega}) to trade $\Omega_d$ into $\Omega_{\mbox{{\scriptsize DDM}}}$ of parent particle if it doesn't decay, by doing so we are able to explore a large hierarchy region of relativistic mass ratio, the exact value of which has little effect as shown in the previous sample calculations. Another way for expression is $\Omega_{\mbox{{\scriptsize DDM}}}h^2=274.2\frac{M_{\mbox{{\tiny DDM}}}}{\mbox{keV}}Y$, where $Y\equiv\frac{n_{\mbox{{\tiny DDM}}}}{s}$ and $n_{\mbox{{\tiny DDM}}}$ and $s$ are separately the DDM number and entropy densities.

The allowed region can be analytically expressed as $(\frac{M_{\mbox{{\tiny DDM}}}}{\mbox{{\scriptsize keV}}}Y)^2\frac{T_d}{\mbox{{\scriptsize yr}}}\lesssim5\times10^{-5}$ at 95\% confidence level, which is shown in Fig.\ \ref{relativisticplot}. Thanks to the high precision WMAP data, it improves the previous one by 3 orders of magnitude.

\section{Discussion}

We see no supporting evidence to introduce the DDM, minimal $\Lambda$CDM is still among the best fit. We are not using the small scale such as the halo structure data, so we do not face the small scale structure problem of \cite{Cembranos:2005us}. Here the scale of structure discrepancy is further constrained to be below 0.5Mpc.

From high energy physics model building perspective, our results put interesting constraint on the gauge mediated supersymmetry breaking (GMSB) scenario in general, in which gravitino is usually a very light LSP. Generically in that kind of model the gravitino mass is in hierarchy with all the other LSP candidate, and combined with large reheating temperature after inflation it is overproduced, probably by many orders. Our results covers all the gravitino related decay process which is suppressed by the Planck scale, where gravitino is either the parent particle or one of the daughter particle. It is very hard to introduce a decay process which can solve the so called ``gravitino problem'' and does not contradicts with observation, and the constraint is done in the most inevitable way, that the process can even escape electromagnetic and hadronic constraints.

Going beyond the gravitino model in GMSB, this constraint can still judge other exotic models which satisfy collisionless daughter particles condition. Moreover, we would argue that this pure gravitational bound should apply to any DDM model even with interactions other than gravitation, since gravitational effect should always be the weakest.

\section*{Acknowledgments}
RH wish to thank Carlos E.~M.~Wagner, Wayne Hu, Tower Wang, Jonathan Feng, Haibo Yu for useful discussion and information. RH is supported partially by Robert R.\ McCormick Fellowship of the University of Chicago.

\appendix

  C.~L.~Reichardt {\it et al.},
  Astrophys.\ J.\  {\bf 694}, 1200 (2009)
  [arXiv:0801.1491 [astro-ph]].


  J.~L.~Sievers {\it et al.},
  arXiv:0901.4540 [astro-ph.CO].


  W.~C.~Jones {\it et al.},
  Astrophys.\ J.\  {\bf 647}, 823 (2006)
  [arXiv:astro-ph/0507494].


\bibitem{Reid:2009xm}
  B.~A.~Reid {\it et al.},
  Mon.\ Not.\ Roy.\ Astron.\ Soc.\  {\bf 404}, 60 (2010)
  [arXiv:0907.1659 [astro-ph.CO]].


  M.~Tegmark {\it et al.}  [SDSS Collaboration],
  Phys.\ Rev.\  D {\bf 74}, 123507 (2006)
  [arXiv:astro-ph/0608632].


  M.~Colless {\it et al.},
  [arXiv:astro-ph/0306581].


\end{document}